\documentclass[aps,prl,showpacs,twocolumn,10pt]{revtex4-1}
\usepackage{amsmath}
\usepackage{amssymb}
\usepackage{dsfont}
\usepackage{graphicx}
\newcommand{\half}{\mbox{$\textstyle \frac{1}{2}$}}

\begin{document}
\title[Quaternionic quantum mechanics and space-time]{Six-dimensional 
space-time from quaternionic quantum mechanics}

\author{Dorje C. Brody${}^1$ and Eva-Maria Graefe${}^2$}
\affiliation{${}^1$ Mathematical Sciences, Brunel University, Uxbridge UB8 3PH, UK\\ 
${}^2$ Department of Mathematics, Imperial College London, London SW7 2AZ, UK}

\begin{abstract} 
Quaternionic quantum Hamiltonians describing nonrelativistic spin particles require 
the ambient physical space to have five dimensions. The quantum dynamics of a 
spin-$\frac{1}{2}$ particle system characterised by a generic such Hamiltonian is 
worked out in detail. It is shown that there exists, within the structure of quaternionic 
quantum mechanics, a canonical reduction to three spatial dimensions upon which 
standard quantum theory is retrieved. In this dimensional reduction, three of the five 
dynamical variables are shown to oscillate around a cylinder, thus behaving in a quasi 
one-dimensional manner at large distances. An analogous mechanism is shown to 
exist in the case of octavic Hamiltonians, where the ambient physical space has 
nine dimensions. Possible experimental tests in search for the signature of extra 
dimensions at low energies are briefly discussed. 
\end{abstract}

\maketitle
\vspace{0.4cm}


In many models that attempt to reconcile quantum theory with gravity, the notion 
of extra dimensions is introduced. If we take seriously the hypothesis that these 
extra dimensions may be relevant to physical reality, then we should likewise take 
seriously the quantum theory underlying these models. Yet, surprisingly little 
attention has been paid to foundational investigations into measurable effects of 
higher-dimensional quantum mechanics. It is well known that extra dimensions can 
change spectral properties of particles, but the standard argument is that if the sizes 
of extra dimensions are sufficiently small, then low energy spectra are typically 
unaltered, and indications of the existence of extra dimensions may be revealed 
only at inaccessibly large energies~\cite{zwiebach}. Quantised energy spectra of 
particles, however, are not the only quantum effect measured in laboratories. It 
appears that other quantum effects arising, e.g., from geometric phase, interference, 
or entanglement, that may be used to probe extra dimensions at low energies, have 
not been fully explored. Furthermore, the following issue concerning 
higher-dimensional quantum theory is often overlooked: The spin-orbit interaction in 
standard quantum mechanics naturally singles out four-dimensional space-time. 
There seems to be no structure, within the complex framework, that allows 
for higher dimensional extensions. 

Here we take the first step towards addressing these fundamental issues and 
exploring the possibility of detecting higher-dimensional quantum effects at low 
energies by investigating certain quaternionic extensions of quantum mechanics 
that naturally lead to six-dimensional space-time structures. Specifically, we 
analyse  the dynamical aspects of a two-level system in quaternionic quantum 
mechanics. Two-level systems are of great importance in many physical 
applications, both as approximations in cases where only two states are of 
relevance to the dynamics, and in the description of the internal degrees of 
freedom for spin-$\frac{1}{2}$ particles. We show that there is an intrinsic 
mechanism for dimensional reduction such that observed phenomena in three 
spatial dimensions can be restored. Similarly, octavic quantum mechanics is 
shown to lead to nine spatial dimensions---a dimensionality often considered in 
string theory models---within which three-dimensional space is naturally embedded. 
By determining dynamical aspects of quaternionic and octavic quantum states of a 
spin particle, we point out the way towards the possible detection of extra dimensions 
at low energies. 

We begin by remarking that there are two fundamental ways in which the use 
of quaternions in physics is related to the notion of six-dimensional space-time. 
The first is the representation of space-time points in terms of quaternionic spinors: 
The points of four-dimensional Minkowski space correspond to two-by-two 
Hermitian matrices $\{x^{AA'}\}_{A,A'=1,2}$. Lorentz transformations are given by 
conjugating $x^{AA'}$ by elements of SL(2,${\mathds C}$), and the Minkowski 
metric for the interval between two points is given by the determinant of their 
difference \cite{Penrose,BH}. In a standard basis this correspondence reads
\begin{eqnarray}
x^{AA'}=\frac{1}{\sqrt{2}}\left( \begin{array}{cc} t+z & x-iy \\ x+iy & t-z 
\end{array} \right) \longleftrightarrow (t,x,y,z) , \label{eq:1}
\end{eqnarray}
and we have the relation $2\det(x^{AA'}) = t^2-x^2-y^2-z^2$. 
Similarly, points of six-dimensional Minkowski space correspond to two-by-two 
quaternionic Hermitian matrices of the form (\ref{eq:1}) with $i$ replaced by 
${\boldsymbol i} = (iy_1+jy_2+ky_3)/y$, where $y^2=y_1^2+y_2^2+y_3^2$. 
Here, $i$, $j$, and $k$ denote three imaginary units of a quaternion, satisfying 
$i^2 = j^2 = k^2 = ijk = -1$ and the cyclic relations $ij=-ji=k$, $jk=-kj=i$, $ki=-ik=j$. 
In this case we have the relation $2\det(x^{AA'}) = t^2-x^2-y_1^2-
y_2^2-y_3^2-z^2$. These two correspondences are related to the 
facts that the universal covering group of SO(3,1) is isomorphic to 
SL(2,${\mathds C}$), and  that of SO(5,1) is isomorphic to 
SL(2,${\mathds H}$), where ${\mathds H}$ denotes the field of quaternions 
\cite{Kugo}. 

Perhaps what is less appreciated is the second connection between quaternions 
and six dimensions arising in the context of quantum mechanics. Complex 
Hermitian matrices represent physical observables in conventional quantum 
mechanics. A trace-free two-by-two complex Hermitian 
matrix, for instance, represents the energy of a spin-$\frac{1}{2}$ particle. The 
spin-orbit interaction of elementary quantum mechanics then requires that the 
(Euclidean) space-time dimension is four. Mathematically, this can be 
seen from the fact that the state space ${\mathds C}{\mathbb P}^1\simeq S^2$ 
of a spin-$\frac{1}{2}$ particle system, obtained by the identification 
$|{\mathit\Psi}\rangle\sim\lambda|{\mathit\Psi}\rangle$, 
$\lambda\in{\mathds C}-\{0\}$, admits 
a natural embedding in ${\mathds R}^3$, and this allows us to make the 
so-called Pauli correspondence whereby we can speak of ``spin in such 
and such direction''. The group isomorphism that underlies this identification 
is that between the universal covering group Spin(3) of SO(3) and the 
two-by-two complex unitary matrices SU(2)$\,\simeq\,$Sp(1). 

Similarly, we can regard a trace-free two-by-two quaternionic Hermitian 
matrix representing the energy of a spin-$\frac{1}{2}$ particle in quaternionic 
quantum mechanics. Then the spin-orbit interaction demands that the 
(Euclidean) space-time dimension is six \cite{BG} (see also \cite{Arnold}). Here 
the Pauli correspondence 
is characterised by the fact that the state space ${\mathds H}{\mathbb P}^1
\simeq S^4$ of a spin-$\frac{1}{2}$ particle system, obtained by the identification 
$|{\mathit\Psi}\rangle\sim|{\mathit\Psi}\rangle\lambda$, 
$\lambda\in{\mathds H}-\{0\}$, admits a natural embedding in 
${\mathds R}^5$. Alternatively stated, there is an isomorphism between the 
universal covering group Spin(5) of SO(5) and the group of 
two-by-two quaternionic unitary matrices Sp(2). (A third connection 
between quaternions and six-dimensional cosmology has been noted by 
Dirac~\cite{Dirac1945}.) We thus see that, be it Euclidean or Lorenzian, 
complex Hermitian form naturally leads to the notion of four-dimensional 
space-time, and quaternionic Hermitian form naturally leads to the notion of 
six-dimensional space-time. Evidently, octavic Hermitian forms lead to 
dimensionality ten. 

The quaternionic Schr\"odinger equation
\begin{eqnarray}
|{\dot{\mathit\Psi}}\rangle = - {\boldsymbol i} {\hat H} |{\mathit\Psi}\rangle, 
\label{eq:2} 
\end{eqnarray}
with ${\hat H}$ Hermitian and ${\boldsymbol i}$ skew-Hermitian unitary, 
generates a unitary time evolution if both ${\hat H}$ and ${\boldsymbol i}$ 
commute with  ${\hat U}_t=\exp(-{\boldsymbol i}{\hat H}t)$. One standard 
approach is to regard ${\boldsymbol i}{\hat H}$ as a generic skew-Hermitian 
operator \cite{Adler}. Another approach, which we shall follow here, is to 
impose a superselection rule that fixes ${\boldsymbol i}$ and restrict ${\hat H}$ 
to the ones that commute with ${\boldsymbol i}$ \cite{Finkelstein}. The 
condition $[{\boldsymbol i},{\hat H}]=0$ thus implies that the specification 
of the Hamiltonian \textit{a fortiori} determines the superselection rule dynamically. 

For a two-level system, a generic quaternionic Hermitian Hamiltonian 
can be expressed in the form 
\begin{eqnarray}
{\hat H} = u_0 {\mathds 1} + \sum_{l=1}^5 u_l {\hat\sigma}_l, 
\label{eq:3}
\end{eqnarray}
where $\{u_l\}_{l=0\ldots 5}\in{\mathds R}$, and 
\begin{eqnarray}
{\hat\sigma}_1 &=& \left( \begin{array}{cc} 0 & 1 \\ 
1 & 0 \end{array} \right), \quad\! 
{\hat\sigma}_2 = \left( \begin{array}{cc} 0 & -i \\ 
i & 0 \end{array} \right), \quad\!
{\hat\sigma}_3 = \left( \begin{array}{cc} 1 & 0 \\ 
0 & -1 \end{array} \right), \nonumber \\ && 
{\hat\sigma}_4 = \left( \begin{array}{cc} 0 & -j \\ 
j & 0 \end{array} \right), \quad\! 
{\hat\sigma}_5 = \left( \begin{array}{cc} 0 & -k \\ 
k & 0 \end{array} \right) \label{eq:4}
\end{eqnarray} 
are the quaternionic Pauli matrices. This follows from the fact that elements of 
a quaternionic Hermitian matrix satisfy $H_{mn}={\bar H}_{nm}$. Then the right 
eigenvalues $E_\pm$ of ${\hat H}$ in (\ref{eq:3}), determined by 
${\hat H}|\phi_\pm\rangle=|\phi_\pm\rangle E_\pm$, are real. Having specified 
the Hamiltonian (\ref{eq:3}) we must select a unit imaginary quaternion such 
that the evolution operator ${\hat U}_t=\exp(-{\boldsymbol i}{\hat H}t)$ is unitary. 
This is given by 
\begin{eqnarray}
{\boldsymbol i} = (iu_2+ju_4+ku_5)/\nu, \label{eq:5}
\end{eqnarray}
where $\nu=\sqrt{u_2^2+u_4^2+u_5^2}$. Then the Schr\"odinger equation 
(\ref{eq:2}) can be expressed more explicitly in terms of the components 
$(\psi_1,\psi_2)$ of the state vector $|{\mathit\Psi}\rangle$ as follows: 
\begin{eqnarray}
\left( \begin{array}{c} {\dot\psi}_1 \\ {\dot\psi}_2 \end{array} \right) = 
\left( \begin{array}{c} 
-(u_0+u_3) {\boldsymbol i} \psi_1 - u_1 {\boldsymbol i} \psi_2 - \nu \psi_2 \\ 
-(u_0-u_3) {\boldsymbol i} \psi_2 - u_1 {\boldsymbol i} \psi_1 + \nu \psi_1 
\end{array} \right) .
\label{eq:6}
\end{eqnarray} 

We can think of the Hamiltonian (\ref{eq:3}) as representing the interaction of 
a `spin vector' ${\vec\sigma}$ with an external field 
${\vec B}=(u_1,u_2,u_3,u_4,u_5)$ in five dimensions. The quaternionic Pauli 
matrices are related to the ten generators of the rotation group SO(5), in a way 
similar to the relation between the three Pauli matrices and the group SO(3). 
The ten skew-Hermitian generators ${\hat\Sigma}_{mn}=\frac{1}{2}[
{\hat\sigma}_m,{\hat\sigma}_n]$ of the dynamics, each inducing a rotation 
that mixes ${\hat\sigma}_m$ and ${\hat\sigma}_n$, fulfil the 
algebraic relation $[{\hat\Sigma}_{mn},{\hat\Sigma}_{m'n'}] = \delta_{mm'}
{\hat\Sigma}_{nn'} + \delta_{nn'}{\hat\Sigma}_{mm'} - \delta_{mn'}
{\hat\Sigma}_{nm'} - \delta_{nm'}{\hat\Sigma}_{mn'}$. The spin 
vector can be seen to fulfil formally the `superspin' algebra of 
Zhang~\cite{Zhang}: $[{\hat\Sigma}_{lm},{\hat\sigma}_n]=\delta_{mn}
{\hat\sigma}_l-\delta_{ln}{\hat\sigma}_m$. The generator of the evolution 
operator is then expressed 
\begin{eqnarray}
{\boldsymbol i}{\hat H} &=& \nu{\hat\Sigma}_{31} + u_0
(u_2{\hat\Sigma}_{54}+u_4{\hat\Sigma}_{25}+u_5{\hat\Sigma}_{42})/\nu  
\nonumber \\ 
&& + u_1 (u_2{\hat\Sigma}_{23} +u_4{\hat\Sigma}_{43}+u_5 
{\hat\Sigma}_{53})/\nu \label{eq:7} \\ 
&& + u_3 (u_2{\hat\Sigma}_{12} +u_4{\hat\Sigma}_{14}+u_5 
{\hat\Sigma}_{15})/\nu  .
\nonumber 
\end{eqnarray}
We see that while each of the ten generators of pairwise-mixing rotations appear 
once, there are only six degrees of freedom. This follows from the Hermiticity 
condition imposed on $\hat H$. The time evolution thus gives rise to certain rotations 
in five-dimensional space.

To determine the dynamics we introduce a quaternionic Bloch vector 
${\vec\sigma}$, whose components are given by 
\begin{eqnarray}
\sigma_l = \langle{\mathit\Psi}|{\hat\sigma}_l|{\mathit\Psi}\rangle/
\langle{\mathit\Psi}|{\mathit\Psi}\rangle, \quad l=1,\ldots,5. \label{eq:8}
\end{eqnarray}
Then for each component we work out the dynamics by making use of the 
Schr\"odinger equation (\ref{eq:6}). After rearrangements we deduce that 
\begin{eqnarray}
\half {\dot\sigma}_1 &=& \nu \sigma_3 - u_3 (u_2\sigma_2 
+ u_4\sigma_4 + u_5 \sigma_5)/\nu \nonumber \\ 
\half {\dot\sigma}_2 &=&  (u_2u_3\sigma_1 - u_1u_2 \sigma_3 
+ u_0u_5\sigma_4 - u_0u_4 \sigma_5)/\nu \nonumber \\ 
\half {\dot\sigma}_3 &=& -\nu \sigma_1 + u_1 (u_2\sigma_2 
+ u_4\sigma_4 + u_5 \sigma_5)/\nu \label{eq:9} \\ 
\half {\dot\sigma}_4 &=& (u_3u_4\sigma_1 - u_0u_5 \sigma_2 
- u_1u_4\sigma_3 + u_0u_2 \sigma_5)/\nu \nonumber \\ 
\half {\dot\sigma}_5 &=& (u_3u_5\sigma_1 + u_0u_4 \sigma_2 
- u_1u_5\sigma_3 - u_0u_2 \sigma_4)/\nu. \nonumber 
\end{eqnarray}
These equations constitute the general quaternionic Bloch equations. The 
special case of (\ref{eq:9}) for which $u_1=\cdots=u_5=0$, i.e. when ${\hat H}=
u_0{\mathds 1}$, has previously been obtained by Wolff \cite{Wolff}. These 
evolution equations preserve the normalisation condition: 
\begin{eqnarray}
\sigma_1^2 + \sigma_2^2 + \sigma_3^2 + \sigma_4^2 + \sigma_5^2 = 1, 
\label{eq:10}
\end{eqnarray}
which can be interpreted as the defining equation for the state space $S^4$. 

As in any physical theory modelled on a higher-dimensional space-time, it is 
important to identify a dimensional reduction leading to a theory consistent with 
observed phenomena perceived in three 
spatial dimensions. In the present context, this amounts to finding a reduction 
of the dynamics on $S^4$ to the conventional Bloch sphere $S^2$. For this 
purpose, let us define the three spin variables according to 
\begin{eqnarray}
\!\!\!\! \sigma_x = \sigma_1, \quad \!\!\!\! \sigma_y = (u_2\sigma_2
+ u_4\sigma_4+u_5\sigma_5)/\nu, \quad \!\!\!\! \sigma_z = \sigma_3. 
\label{eq:11}
\end{eqnarray}
Then it follows from (\ref{eq:9}) that
\begin{eqnarray}
\half {\dot\sigma}_x &=& \nu \sigma_z -u_3\sigma_y \nonumber \\ 
\half {\dot\sigma}_y &=& u_3\sigma_x - u_1\sigma_z \label{eq:12} \\ 
\half {\dot\sigma}_z &=& u_1 \sigma_y - \nu \sigma_x.  \nonumber 
\end{eqnarray} 
These equations are, indeed, the standard Bloch equations for a 
spin-$\frac{1}{2}$ particle immersed in a magnetic field with strength 
${\vec B}=(u_1,\nu,u_3)$. The reduced spin dynamics is thus confined to the 
state space 
\begin{eqnarray}
\sigma_x^2+\sigma_y^2+\sigma_z^2=r^2, \label{eq:13}
\end{eqnarray}
where $r\leq1$ is time independent. The dynamical equations (\ref{eq:12}) thus 
generate Rabi oscillations on the reduced state space $S^2$ about the axis 
$(u_1,\nu,u_3)$, with angular frequency $\omega$, where $\omega^2=4(u_1^2+
u_2^2+u_3^2+u_4^2+u_5^2)$. 

\begin{figure}[t]
\begin{center}
  \includegraphics[width=7.5cm]{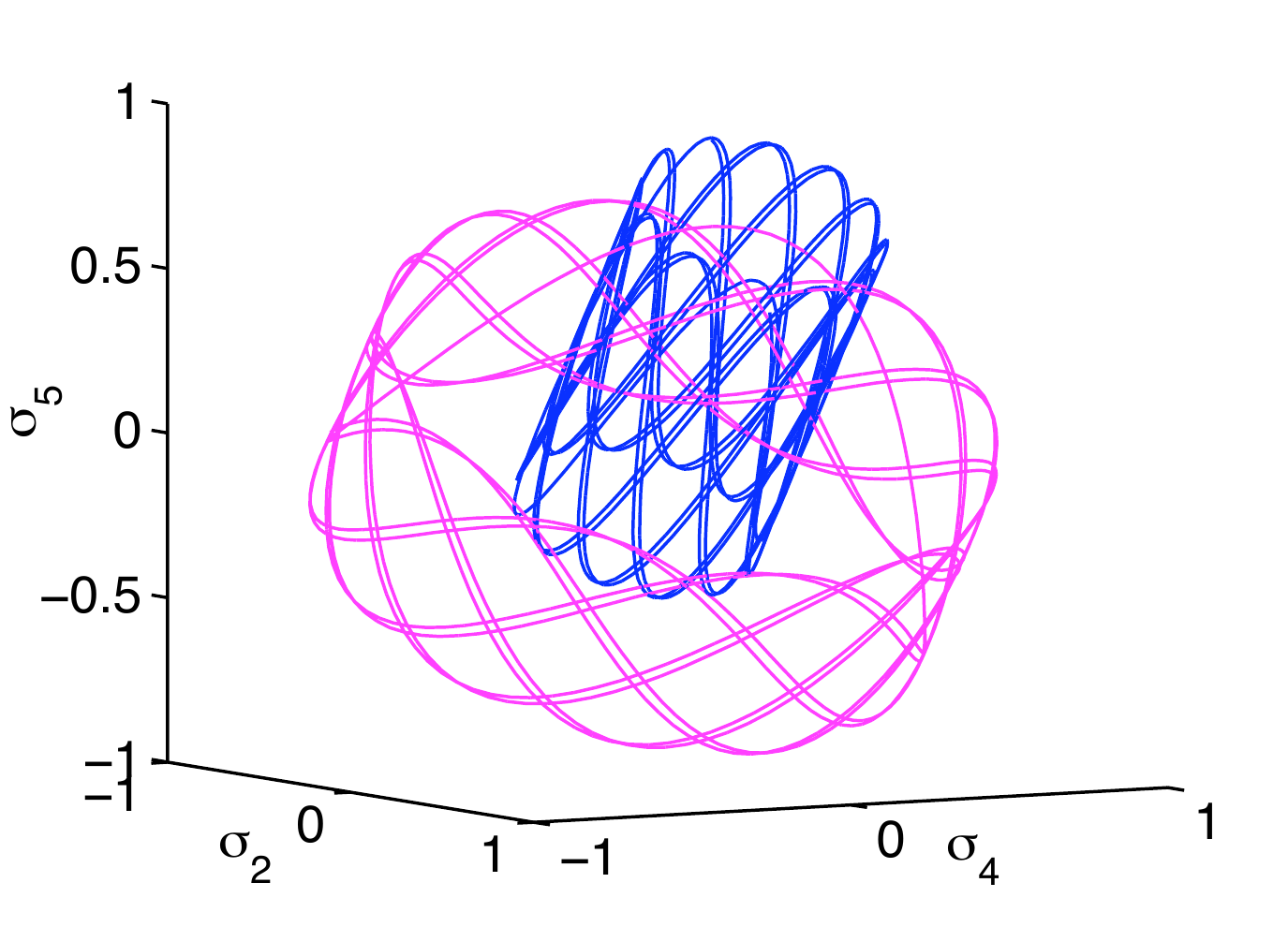}
  \caption{(colour online) 
 Examples of dynamical trajectories traced by the variables 
  $(\sigma_2(t),\sigma_4(t), \sigma_5(t))$ for different initial conditions. For each 
  choice of $c$ the orbits form cylindrical Rabi oscillations. The axis of the 
  cylinder is determined by the vector $(u_2,u_4,u_5)$. 
  \label{fig:1} 
  }
\end{center}
\end{figure}

To identify the structure characterising the evolution of the `internal' dynamical 
variables of $\sigma_y$: $\sigma_2$, $\sigma_4$, and $\sigma_5$, let us 
subtract (\ref{eq:13}) from (\ref{eq:10}) to eliminate $\sigma_1$ and $\sigma_3$. 
Then we deduce that the motion lies on a cylinder in ${\mathds R}^3$: 
\begin{eqnarray}
(u_2\sigma_4-u_4\sigma_2)^2 &+& (u_4\sigma_5-u_5\sigma_4)^2 \nonumber 
\\ &+& (u_5\sigma_2-u_2\sigma_5)^2 = \nu^2 c^2, \label{eq:14} 
\end{eqnarray}
that is, $|(u_2,u_4,u_5)\times(\sigma_2,\sigma_4,\sigma_5)|=\nu c$, 
where $c^2=1-r^2$ is the squared radius of the cylinder, 
whose axis points in the $y$-direction. In figure~\ref{fig:1} we plot typical 
motions of the variables $\sigma_2$, $\sigma_4$, $\sigma_5$ on the cylinder. 

The time evolution of these dynamical variables can also be 
represented in the form of Bloch equations if we transform to the auxiliary 
variables $\sigma_{y_1}=u_4\sigma_5-u_5\sigma_4$, $\sigma_{y_2}=
u_5\sigma_2-u_2\sigma_5$, and $\sigma_{y_3}=u_2\sigma_4-u_4\sigma_2$. 
Then we have ${\dot\sigma}_{y_1} = 2u_0 (u_5\sigma_{y_2} -u_4
\sigma_{y_3})/\nu$, ${\dot\sigma}_{y_2} = 2u_0 (u_2\sigma_{y_3} - u_5
\sigma_{y_1})/\nu$, and ${\dot\sigma}_{y_3} = 2u_0 (u_4 \sigma_{y_1} - u_2 
\sigma_{y_2})/\nu$. 
These variables are useful in understanding the dynamics in five dimensions: 
We let ${\hat\sigma}_{x,y,z}$ be the operators for $\sigma_{x,y,z}$, and 
${\hat\sigma}_{y_1,y_2,y_3}$ be the operators for ${\sigma}_{y_1,y_2,y_3}$. 
Additionally, define a new set of rotation generators by 
${\hat\Sigma}_x=\frac{1}{2}[{\hat\sigma}_y,{\hat\sigma}_z]$, 
${\hat\Sigma}_y=\frac{1}{2}[{\hat\sigma}_z,{\hat\sigma}_x]$, 
${\hat\Sigma}_z=\frac{1}{2}[{\hat\sigma}_x,{\hat\sigma}_y]$, 
${\hat\Sigma}_{y_1}={\hat\Sigma}_{54}$,   
${\hat\Sigma}_{y_2}={\hat\Sigma}_{25}$, and 
${\hat\Sigma}_{y_3}={\hat\Sigma}_{42}$. These operators fulfil a pair of closed 
algebraic relations $\frac{1}{2}[{\hat\Sigma}_a,{\hat\Sigma}_b]=-\epsilon_{abc}
{\hat\Sigma}_c$ for $a,b,c$ ranging over $x,y,z$; and 
$\frac{1}{2}[{\hat\Sigma}_{y_l},{\hat\Sigma}_{y_m}]=\epsilon_{lmn}
{\hat\Sigma}_{y_n}$ for $l,m,n$ ranging over $1,2,3$. Then (\ref{eq:7}) can 
be expressed in the concise form 
\begin{eqnarray}
{\boldsymbol i}{\hat H} = u_1{\hat\Sigma}_{x} + \nu{\hat\Sigma}_{y}
+u_3{\hat\Sigma}_{z}  + u_0 {\hat\Sigma}_\perp, \label{eq:16} 
\end{eqnarray} 
where ${\hat\Sigma}_\perp=(u_2{\hat\Sigma}_{y_1} +u_4{\hat\Sigma}_{y_2}+
u_5{\hat\Sigma}_{y_3})/\nu$ is the generator of the planar rotation about the 
three-space spanned by $x,y,z$. In this manner we see how the subgroup 
SO(3)$\times$U(1) of SO(5) emerges naturally, on account of the fact 
that $[{\hat\Sigma}_{x,y,z}, {\hat\Sigma}_\perp]=0$. In particular, if 
${\rm tr}\,{\boldsymbol i}{\hat H}=0$, i.e. if $u_0=0$, then it is not possible to 
detect extra dimensions dynamically. 

This result shows that the superselection rule for ${\boldsymbol i}$ emerges 
from symmetry breaking. In complex quantum mechanics, given a state one 
can always unitarily transform it to another arbitrary state by a suitable choice 
of Hamiltonian. In quaternionic quantum mechanics with the superselection 
rule (\ref{eq:5}), the ratio $u_2:u_4:u_5$ is fixed so that the only parametric 
freedom in the Hamiltonian are those appearing in (\ref{eq:16}). It follows that 
a state with a given value of $r$ in (\ref{eq:13}) cannot unitarily evolve into 
another state with a different value of $r$.

The superselection rule resulting from the symmetry breaking circumvents 
a difficulty associated with combined systems in quaternionic quantum 
mechanics (cf.~\cite{Adler,Horwitz2,Baez}). If all systems share the same 
${\boldsymbol i}$, then one is working with a commuting subalgebra of 
quaternions; thus circumventing the issues associated with the construction 
of tensor products for combined systems. While the standard choice of 
complex quantum mechanics ${\boldsymbol i}=i$ can be regarded as a 
special case of this formalism, the embedding into the quaternionic space 
nevertheless accommodates extra dimensions. These extra dimensions 
are not introduced `by hand'; rather, they emerge from the requirement of 
unitary time evolution generated by a Hermitian quaternionic Hamiltonian of 
a two-level system. Furthermore, the resulting dynamics naturally 
factorises into a motion in a three-space and a motion for the remaining 
`hidden coordinates'. 

It is worth remarking that the structure revealed in the foregoing analysis carries 
through to an octavic representation of a spin-$\frac{1}{2}$ system. In this case, 
the spin vector ${\vec\sigma}$ lies on an eight sphere $S^8\subset{\mathds R}^9$. 
If we define $\sigma_y$ in a manner analogous to (\ref{eq:11}) involving the seven 
spin components $\sigma_2, \sigma_4,\cdots,\sigma_9$, then a calculation shows 
that the dynamical equations satisfied by the reduced spin variables are given by 
(\ref{eq:12}), with $\nu^2=u_2^2+u_4^2+\cdots+u_9^2$. To characterise the surface 
upon which the remaining degrees of freedom are confined, let us write 
$[l,m,n] = |(u_l,u_m,u_n)\times(\sigma_l,\sigma_m,\sigma_n)|^2$. 
Hence the left side of (\ref{eq:14}), for instance, 
becomes $[2,4,5]$. Then in the octavic case these dynamical variables are 
confined to a real six-dimensional manifold determined by the relation: 
\begin{eqnarray}
[2,4,5] &+& [2,6,7]+[2,8,9]+[4,6,8] \nonumber \\ &+& 
[4,7,9]+[5,6,9]+[5,7,8]=\nu^2c^2. \label{eq:17} 
\end{eqnarray} 
This manifold, which is the octavic generalisation of (\ref{eq:14}), has the 
structure of a cylinder $S^5\times{\mathds R}^1$ in the direction of the vector 
$(u_2,u_4,u_5,u_6,u_7,u_8,u_9)$, with radius $c$. 

It is important to note that here we consider dynamics in the angular momentum 
space, and that the `thickness' $c$ of the $y$-axis is not related to the size of 
extra dimensions in coordinate space. The higher-dimensional angular momentum 
discussed here can be related to a higher-dimensional coordinate space in the 
usual manner: $L_{mn}=x_m p_n -p_n x_m$, with $p_n={\boldsymbol i}\partial_n$. 
The size of the $x_n$ does not affect the size of $c$. 

We conclude by discussing the possibility of detecting extra dimensions in a 
laboratory. An experimental test for quaternionic quantum mechanics has 
previously been propose by Peres \cite{Peres}, which has subsequently been 
shown to yield null outcome by Adler \cite{Adler2}. Given the analysis presented 
here of a quaternionic spin system, another obvious proposal arises from relation 
(\ref{eq:13}), since the left side involves quantities that can be estimated directly 
from experimental data, whereas the value of the right side, according to complex 
quantum mechanics, is unity. However, in the quaternionic case there are states 
for which $c>0$, and we have $r^2=1-c^2<1$. To perform an experiment, one 
prepares a large number of spin-$\frac{1}{2}$ particles in a pure state and measures 
the spin in three orthogonal directions to estimate $\sigma_x^2+
\sigma_y^2+\sigma_z^2$. If the result is less than one, then this gives a strong 
indication that there can be extra dimensions. 
 
Although such a basic experiment is easily performed, it need not constitute 
a useful test for the following two reasons: (i) the prepared states must 
be pure; and (ii) the measurements have to be performed along three 
strictly orthogonal directions. Any impurity or deviation from 
orthogonality will lead to a number less than one even in three dimensions. Hence 
it may be difficult to extract useful insights from this simple experiment. 
Nevertheless, this example illustrates 
the important point that in principle it is possible to probe extra dimensions at low 
energies. Viable experiments may be constructed by making use 
of interference effects arising from, for instance, geometric phases (cf. 
\cite{Anandan,Zhang2,Hasebe}). Alternatively, the existence 
of an SO(5) symmetry between antiferromagnetic and superconducting phases 
that can be described by a five-dimensional superspin \cite{Zhang} might provide 
a clue along this line of investigation; and, conversely, an extension of Zhang's 
SO(5) representation to SO(9) might lead to new predictions in superconductor 
physics. The identifications made here of the structures of `commutative' 
quaternionic and octavic state spaces will undoubtedly help in making progress 
towards these directions. 

\noindent EMG is supported by the Imperial College JRF scheme.

\vspace{-0.6cm}




\begin{thebibliography}{999}

\vspace{-0.4cm}

\bibitem{zwiebach} 
Zwiebach,~B. 2004 
{\em A First Course in String Theory} 
(Cambridge: Cambridge University Press). 

%
\bibitem{Penrose} 
Penrose,~R. \& Rindler,~W. 1984 
{\em Spinors and Space-Time}, vol. 1 
(Cambridge: Cambridge University Press).

\bibitem{BH} 
Brody,~D.~C. \& Hughston,~L.~P. 2005 
Theory of quantum space-time. 
{\em Proc. R. Soc. Lond.} A\textbf{461} 2679. 

\bibitem{Kugo} 
Kugo,~T. \& Townsend,~P. 1983 
Supersymmetry and the division algebras. 
{\em Nucl. Phys.} B\textbf{221}, 357. 

\bibitem{BG} Brody,~D.~C. \& Graefe,~E.~M. 2011 
On complexified mechanics and coquaternions. {\em J. Phys.} 
A\textbf{44}, 072001. 

\bibitem{Arnold} Arnold,~V.~I. 1999 
Symplectization, complexification and mathematical trinities. 
In {\em The Arnoldfest: Proceedings of a conference in honour of V.I. 
Arnold for his sixtieth birthday}. 
Eds. E.~Bierstone \textit{et al}. (Providence, Rhode Island: AMS). 

\bibitem{Dirac1945} 
Dirac,~P.~A.~M. 1945 
Application of quaternions to Lorentz transformations. 
{\em Proc. R. Irish Acad.} A\textbf{50}, 261. 

\bibitem{Adler} Adler,~S.~L. 1995 {\em Quaternionic Quantum 
Mechanics and Quantum Fields} (Oxford: Oxford University 
Press). 

\bibitem{Finkelstein} Finkelstein,~D., Jaueh,~J.~M., Schiminovieh,~S. 
\& Speiser,~D. 1962 Foundations of quaternion quantum mechanics. 
{\em J. Math. Phys.} \textbf{3}, 207. 

\bibitem{Zhang}
Zhang,~S.-C. 1997 
A unified theory based on SO(5) symmetry of superconductivity and 
antiferromagnetism. 
{\em Science} \textbf{275}, 1089. 

\bibitem{Wolff} Wolff,~U. 1981 A quaternion quantum system. 
{\em Phys. Lett.} A\textbf{84}, 89. 


\bibitem{Horwitz2} Razon,~A. \& Horwitz,~L.~P. 1991 
Tensor product of quaternion Hilbert modules. {\em Act. App. Math.} 
\textbf{24}, 141. 

\bibitem{Baez} Baez,~J. 2012 Division algebras and quantum theory. 
{\em Found. Phys.} \textbf{42} (to appear). 

\bibitem{Peres} 
Peres,~A. 1979 
Proposed test for complex versus quaternion quantum theory. 
{\em Phys. Rev. Lett.} \textbf{42}, 683. 

\bibitem{Adler2} Adler,~S.~L. 1988 
Scattering and decay theory for quaternionic quantum mechanics, 
and the structure of induced $T$ nonconservation. 
{\em Phys. Rev.} D\textbf{37}, 3654. 

\bibitem{Anandan}
Adler,~S.~L. \& Anandan,~J. 1996 
Nonadiabatic geometric phase in quaternionic Hilbert space. 
{\em Found. Phys.} \textbf{26}, 1579. 


\bibitem{Zhang2}
Demler,~E. \& Zhang,~S.-C. 1999 
Non-Abelian holonomy of BCS and SDW quasiparticles. 
{\em Ann. Phys.} \textbf{271}, 83. 

\bibitem{Hasebe}
Hasebe,~K. 2005 
Supersymmetric quantum-Hall effect on a fuzzy supersphere. 
{\em Phys. Rev. Lett.} \textbf{94}, 206802. 



\end{thebibliography}
\end{document}